\documentclass[10pt]{article}
\usepackage{epsf}
\usepackage{graphicx}

\title{Fluctuations in some climate parameters}
\author{A.D.Erlykin $^{1,2}$, B.A.Laken $^{3}$ and A.W.Wolfendale $^{1}$\\
$(1)$ Department of Physics, Durham University, Durham, UK\\
$(2)$ Permanent address: P N Lebedev Institute, Moscow, Russia\\
$(3)$ Instituto de Astrofisica de Canarias, Tenerife, Spain}
\begin{document}
\maketitle

\begin{abstract}
There is argument as to the extent to which there has been an increase over the past
 few decades in the frequency of the extremes of climatic parameters, such as 
temperature, storminess, precipitation, etc, an obvious point being that Global Warming
 might be responsible.  Here we report results on those parameters of which we have had
 experience during the last few years : Global surface temperature, Cloud Cover and the
 MODIS Liquid Cloud Fraction. In no case have we found indications that fluctuations of
 these parameters have increased with time.
\end{abstract}

\section{Introduction}
`Global Warming' (GW) has been blamed for a number of the World's ills but, although there 
is agreement that the Global mean temperature is increasing, and with it the frequency 
of extremes of some climate parameters, for others there is not yet a consensus 
(~IPCC, 2007~). For example, the IPCC state that `heat waves will be more intense and 
longer lasting' and `precipitation will increase in tropical, mid- and high-latitude 
regions' but other changes are uncertain. Insofar as such changes might be associated 
with changes in temperature, cloud cover and atmospheric water content - and their 
latitudinal dependence - our work involving 'fluctuations' in these quantities has 
relevance.

There is no doubt about the importance of climate variability over 'short' periods, as 
well as long term 
trends: such changes as can be laid at the door of climate change (~especially droughts
 and floods~) have already had a serious effect on the international economy; for the 
period 1985-1999, the loss to the international Gross World Product was about 0.3\% 
(Houghton, 2009).

It is understandable that GW as such, and its effects, are variable over the Earth's 
surface. Concerning the magnitude of the warming, from say, 1950 to 2000, the increase
for 'sub-continental scale regions' vary from 0.1 to 1.8$^\circ$C, with a rather broad 
distribution (Broham et al., 2006). Such variations might be expected to reflect in 
temporal variations, too.

Concerning the observations of the extremes over the surface of the Earth for 
'storminess', there is some confusion. For the North European and North Atlantic, a 
recent work (Hanna et al, 2008) has found no significant
 increase in storminess since 1830, the period over which good records exist.  However,
 as the authors point out, other workers (e.g. Kaas and Anderson, 1999) have found a 
tendency towards more severe storms in the last few decades in extratropical regions. 
Yet others did not find a sustained longer-termed increase in storminess since the 19th
 Century, (e.g. Alexanderson et al., 1998).

Our own work is restricted in object: to examine the possibility of a changed frequency
 of fluctuations of temperature (GISS.NASA), cloud cover (ISCCP) and liquid cloud 
fraction (MODIS), the 'fluctuations' usually relating to year by year. In some cases 
the period over which the study can be made is rather short, in view of the (satellite)
 data being only of recent availability, specifically, the ISCCP data started to become
 available in 1983. This shortness may not be too serious, however, in view of the 
current rapid rate of surface temperature increase (~0.6$^\circ$C since 1983, compared 
with 0.35$^\circ$C for the previous 60 years, see Figure 1, which comes from the IPCC 
Report, 2007~).

Although the IPCC Report gives a probability of less than 10\% that Global Warming is 
due to causes other than anthropogenic gases (AG) our inspection of the details shows 
that the probability is nearer 2-3\%. In any event we consider that AG are very 
important and since these are generated on Earth in a very non-random manner, in both 
space and time, extremes might be expected to be detected even over a short, recent, 
time base.

Some remarks are necessary about the science behind a link between Global Warming 
and the frequency of extreme 'events'.  A general remark that can be made concerns the 
well-known sensitivity of 'climate' to small changes of input parameters: atmospheric 
and oceanic circulation, precipitation etc. caused by positive feedbacks. An 
illustration of the sensitivity of the atmosphere to perturbations comes from the 
Pinatubo volcano of 1991: the change in oceanic heat was about $3\cdot 10^6$ times the 
energy injected by the volcano itself (~dust injected into stratosphere which reduced 
the intensity of solar radiation reaching the Earth's surface was responsible~). The 
value comes from an estimate of the energy 
involved in the outburst ($\sim 10^{16}J$) compared with the reduction in oceanic 
energy of $\sim 3\cdot 10^{22}J$ (Crouch et al., 2005). A natural consequence of GW is 
that there are local variations, but predictions have not been unanimous 
(Houghton, 2009) and climate modellers are still very active in this area. It is 
because of the expected (~and observed~) local variations, such as those identified by 
the IPCC, that our studies have included the latitude variations of the parameters 
under examination.

We know of no detailed predictions of the likely variability of the parameters to be 
examined by us, although related ones have been made with respect to precipitation. 
Specifically, Alan and Soden (2008) have studied extreme precipitation events in an 
anthropogenically warmed climate. Their obsevations revealed a distinct link between 
extremes of rainfall and temperature, a link stronger than that predicted by climate 
models.  Importantly, work by Tebaldi et al. (2006), also concerning precipitation, has
shown that the standard deviation of the precipitation intensity (~annual total 
precipitation divided by the number of days~) followed the Global surface temperature 
up to the year 2000 (~their limit~) and their prediction for 2000 to 2010 was a marked 
increase as seen. Thus, if this feature is symptomatic of the other climate parameters 
studied by us we might expect an increase in fluctuations.

\section{The expected incidence of climatic fluctuations}

We define the fluctuation as the deviation of the particular value (~e.g. temperature, 
low cloud cover - LCC or liquid cloud fraction - LCF~) from the 6- or 3-degree 
polynomial fit of the temporal behaviour of the parameter in question. It is 
appreciated that all 3 are interconnected, but the fluctuations could well differ.
It is possible that the fluctuations would be different (~bigger?~) for the 
period when, following the anthropogenic effects, the temperature change was bigger.  
It is possible that changes due to variations in solar irradiance, 'SI', alone, which 
many believe was probably responsible for much of the increase in mean Global 
temperature prior to the 1950s (~e.g. Haigh, 2002~) did not 
affect the fluctuations but that the incidence of anthropogenic gases (AG) has done, 
the reason being the difference in the manner of injection of the extra energy: 
external (SI) and internal to the atmosphere (AG), the external being 'smoothly' 
distributed but the latter  'patchily'.

The mean Global surface temperature vs time (Figure 1) can be taken as the starting 
point for an examination of temperature fluctuations. Such fluctuations might be 
expected to start to change in 1950 when AG 
started to bite.  On this basis the fluctuations might increase/decrease continuously
 with time from about 1970, after a transitional period conditioned by thermal capacity
 (~two decades ?~).

\section{Temperature fluctuations}
\subsection{Surface temperature fluctuations vs latitude}

We have used the GISS.NASA temperature series (GISS.NASA) to find the root mean square 
(RMS) temperature vs latitude for the whole time range: 1880 - 2010 with the result 
shown in Figure 2. We are aware of the limitations in the temperature series as a 
function of latitude, particularly in the Polar regions, where the early measurements 
were very sparse, both in place and in time. The relevance of Figure 2 is, in part, to 
search for evidence of the IPCC contention that 'it is very likely that heat waves 
have become more frequent over most land areas' (~over the last 50 years~) and 'it is 
very likely that cold days, cold nights and frosts have become less frequent over most 
land areas, while hot days and hot nights have become more frequent over land'. The 
'land' area aspect can be examined by way of the latitude dependence and this is 
illustrated in Figure 2 where the fraction of land mass (~land/(land+ocean)~)) is also 
plotted as a function of latitude.

Inspection of Figure 2 shows evidence for a general similarity between the dependence 
of RMS temperature on latitude and that of land fraction on latitude. An exception is 
for latitudes above 64$^\circ$N where, although the land fraction is small, the 
temperature fluctuations are large. This divergence is not too unexpected in view of a
 number of 'peculiarities' in this latitude region; most notably the greatest magnitude
 of the '1940's temperature peak' and the undoubted vagaries of polar ice with its 
attendant albedo consequences (~snow changing to water causes a reduced albedo and 
increase SI absorption.

Our study showing the general correlation of temperature fluctuations with land 
fraction corroborates the IPCC conclusions mentioned above. A ready explanation lies in
 the small thermal inertia of the land compared with the oceans allowing land 
temperatures to respond more readily to external forcing.

\subsection{Surface temperature fluctuations vs time for the Globe}

The surface temperature data for the whole Globe have been divided into 6 sets, each of
 20 years duration, and the frequency distributions of monthly values are as shown in 
Figure 3. '20 years' has been taken as the interval so as to give a 'reasonable' number
 of months (~240~) for the statistical analysis in each period. It has no relevance to 
the 11 year or 22 year solar cycle. The absence of marked changes from one interval to 
the next suggests that it is not too long.

Inspection of Figure 3 shows little variation of the RMS with time (~the next section 
will quantify this~) but, coupled with the increase in Global mean temperature with 
time (~Figure 1~), the number of months per year with absolute, not relative 
temperature greater than some particular value has increased markedly. Comparing the 
periods 1910-1929 and
1990-2009 we note that whereas, in the latter, half the months had mean temperatures 
above 14.4$^\circ$C, for the former there were none. This rather obvious result is not 
inconsistent with the IPCC's observation that the frequency of heat waves is 
increasing.

The relative variability of Global mean temperature is perhaps of equally important 
significance in that it bears on the stability of the atmosphere as a function of time.
 Figure 4 shows the RMS deviation vs. time. The variation is interesting in that, if 
anything, it illustrates an increased stability of the temperature since 1930s. Why 
there should have been a significant reduction at this time, which predates the 1970,
 when we had expected changes to occur, is not at all understood.  

\section{Fluctuations in Cloud Cover vs time}

A considerable number of analyses have taken place of the correlations of cloud cover 
with cosmic ray (CR) and solar irradiance (SI) variations.  Particular attention has 
focussed on the low cloud cover (LCC; ISCPP), i.e. those clouds with tops below 3.2 km.

In our previous work (Erlykin and Wolfendale, 2010) we pointed out that the mean LCC 
fell slowly with time, although we considered that it was an artifact due to the 
dependence of mean cloud height on surface temperature, which has been increasing.  
This change would not be expected to have had an effect on the fluctuations in the LCC.

A study has been made of the manner in which the RMS fluctuation in LCC has changed 
with time (~over the whole period~), represented by $b_{LCC-time}(\%y^{-1})$, as a 
function of latitude (~compared with Figure 2 for the RMS temperature for the whole 
period, also as a function of latitude~). The result is shown in Figure 5a. It is 
apparent that there is very little departure from constancy, i.e. cloud fluctuations 
have not varied much with time and are largely independent of latitude.

Another indicator of cloud cover is that provided by MODIS: the `liquid cloud 
fraction' (LCF). Figure 5b shows a comparison of the values for the slopes for all 
latitude bands. Again all slopes are either consistent with zero or negative, which 
means that there is no increase of LCF fluctuations with time.

Interestingly, the mean LCF (~not fluctuation but actual value~) falls with time, 
having a mean linear slope of -0.173$\pm$0.013 \%y$^{-1}$. This can be compared with 
the slope for the total cloud cover (~HCC+MCC+LCC~) of -0.054$\pm$0.010 \%y$^{-1}$
(~from Erlykin and Wolfendale, 2010~). The annual variations can be used as a datum;
 they are (~peak-to-peak~) $\sim$7\% for LCF and $\sim$2.3\% for the total cloud cover.
The ratio of the slopes, for LCF and total cloud cover, is thus about the same as that 
for their respective annual variations. This adds weight to the use of the LCF as well 
as the total cloud cover for our fluctuation studies.
  
\section{Conclusions}

Inspection of the Figures shows three main features, as follows:
\begin{itemize}
\item[(a)] The dependence of fluctuations of annual temperature on latitude is as 
expected in the sense that (~with the exception of the anomalous North Polar region~) 
it correlates well with land fraction: latitude bands with higher oceanic fractions 
having smaller fluctuations. This is a natural consequence of the oceans having a 
higher thermal capacity than land.
\item[(b)] Cloud fluctuations have not changed with time, at least over the last 20 
years.
\item[(c)] The RMS of the mean monthly Global temperature appears to have fallen with 
time but perhaps in a stepwise fashion. The step occured in the vicinity of 1930, 
several decades before it would have been expected. A conceivable explanation is that 
it represents an early manifestation of the effect of increased anthropogenic gases and
 that some aspects of climate became more stable than before but a more likely cause is
 the inevitable greater uncertainties in the measurements then.

It is unlikely that volcanoes have contributed to the change in RMS: Figure 1 shows the
 major volcanoes in `our' period and it is apparent that they contributed to most of 
the points in Figure 4 so that the effect on the overall trend should be small.
\end{itemize}

The conclusion must be that there is no evidence for an increase in relative 
temperature fluctuations since anthropogenic gases started to bite in or about 1950, at
 least for averages over $20^{\circ}$ latitude ranges. However, there 
needs to be caution because of the problem with the significantly higher 
temperature fluctuations before about 1930.  Going further back than about 1900, - to 
1000 (Crowley, 2000) provides no evidence for a trend in RMS with time over this 
period. Specifically, the mean RMS was $0.08\pm0.02^{\circ}$C for 10-yearly temperature
 values with no significant systematic trend. The values in Figure 4 are higher because
 they refer to the spread of monthly values; the 10-yearly values have considerable 
smoothing.

None of the foregoing is at variance with the conclusion of the IPCC Report, indeed
it is supportive of it in pointing up the importance of distinguishing between oceanic 
and land changes. Concerning 'heat waves', although the fluctuations in temperature 
about their bi-decadal means have not changed (~Figure 3~) the frequency of appreciable
 numbers of months having temperature greater than a particular absolute temperature 
has, of course, increased with time.

{\bf Acknowledgements}

The authors thank the Dr John Taylor Charitable Foundation for financial support.
A reviewer is thanked particularly for advice on a possible interpretation of Figure 4. \\

{\bf References}

\begin{itemize}
\item[1.] Allan, R.P. and Soden, B.J. (2008). 'Atmospheric warming and the 
amplification of precipitation extremes', Science, \textbf{321}, 1481.
\item[2.] Alexanderson, H, Schmith, T, Iden, K and Tuomenvirta, H. (1998). `Long term 
variations of the storm climate over NW Europe'. Global Atmos. Ocean Syst. \textbf{6} 
97.
\item[3.] Broham, P., Kennedy, J.J., Tett, S.F.B. and Jones, P.D. (2006). 
'Uncertainty estimates in regional and global observed temperature changes: a new data 
set from 1850', J. Geophys. Res. \textbf{111}, D12106.
\item[4.] Church, J.A., White, N.J. and Arblaster, J.M. (2005). 'Significant 
decadal scale impact of volcanic eruptions on sea level and ocean heat content', 
Nature, \textbf{438}, 74.
\item[5.] Crowley, J J. (2000) `Causes of Climate Change over the past 1000 y'. 
Science, \textbf{289}, 270.
\item[6.] Erlykin, A.D. and Wolfendale, A.W. (2009). `Global Cloud Cover and the 
Earth's Mean Surface Temperature'. Surv. Geophys., \textbf{31}, 399.
\item[7.] GISS: http://data.giss.nasa.gov/gistemp/tabledata/GLB.Ts.txt
\item[8.] Haigh, J.D., (2003), 'The effects of solar variability on the Earth's 
climate', Phil. Trans. Roy. Soc. London, \textbf{A361}, 95
\item[9.] Hanna, E, Cappelen, J, Allan, R, Jonsson, T, le Blancq, F, Lillington, T and 
Hickey, K. (2008). `New insights into North European and North Atlantic Surface 
Pressure Variability, Storminess and Related Climate Change since 1830', J. Climate, 
\textbf{21}, 6739.
\item[10.] Houghton, J. (2009). 'Global Warming', Cambridge University Press. 
\item[11.] IPCC, Fourth Assessment Climate Change Reports, 2007, \\ 
http://www.ipcc.ch/publications.
\item[12.] ISCCP : the data are obtained from the International Satellite Cloud 
Climatology Project : http://isccp.giss.nasa.gov. maintained by the ISCCP research 
group at NASA Goddard Institute for Space Studies, New York.
\item[13.] Kaas, E and Anderson, U. (1999). `Scenarios for extra-tropical storm and 
wave activity : Methodologies and results'. ECLAT-2 Rep. 3 KNMI, 24.
\item[14.] MODIS : http.//disc.sci.gsfc.nasa.gov./data-holdings/PIP.
\item[15.] Tebaldi, C., Hayhoe, K., Arblaster, J.M. and Meetil, G.A. (2006). 
'Going to the Extremes. An Intercomparison of Model-Simulated Historical and Future 
Changes in Extreme Events', Climate Change, DOI: 10.1007/S10584-006-9051-4.
\end{itemize}

{\bf Captions to the Figures} \\

\noindent \textbf{Figure 1} Global mean surface temperatures over the 20th Century from 
observations (~IPCC, 2007, full line~). Vertical dashed lines indicate the 
eruptions of volcanoes: S - Santa Maria, A - Agung, E - El Chichon, P - Pinatubo. 
Horizontal dotted line - the mean Global temperature during the last century: 
13.86$^\circ$C. \\
\textbf{Figure 2} Fluctuations in annual surface temperature vs latitude (full circles)
. Also shown is the ratio of land to (land + ocean) for the various latitude bands 
(open circles). The temperature RMS values relate to $20^\circ$ intervals in latitude 
as do those for the ratio of land to (land + ocean). The lines just join the points, in
 fact the actual variation will be smooth. \\ 
\textbf{Figure 3} Distribution of Global Surface temperatures about the 3-degree 
polynomial fits. \\
\textbf{Figure 4} Temporal variation of the global temperature fluctuations in the last
 130 years. The bracketed value under each point is the difference in average 
temperature for each \textbf{20-year} time interval from the datum level in Figure 1.\\
\textbf{Figure 5} Temporal dependence of the LCC and LCF fluctuations as a function 
of latitude.

\newpage
\begin{figure}[ht]
\begin{center}
\includegraphics[height=15cm,width=12cm,angle=-90]{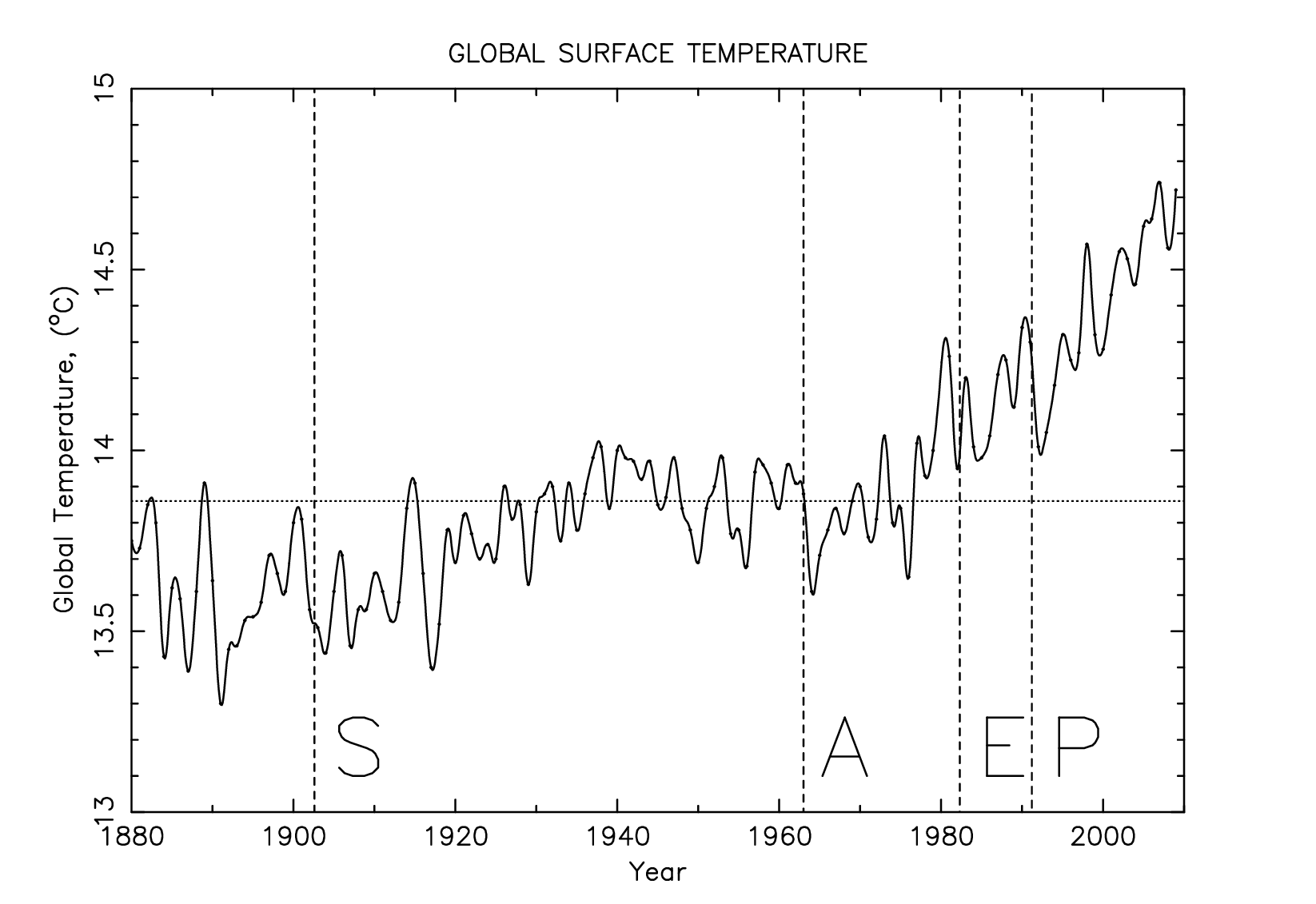}
\caption{}
\label{fig01}
\end{center}
\end{figure}

\newpage
\begin{figure}[ht]
\begin{center}
\includegraphics[height=15cm,width=15cm,angle=-90]{fluc_fig02.eps}
\caption{}
\label{fig02}
\end{center}
\end{figure}

\newpage
\begin{figure}[ht]
\begin{center}
\includegraphics[height=15cm,width=15cm]{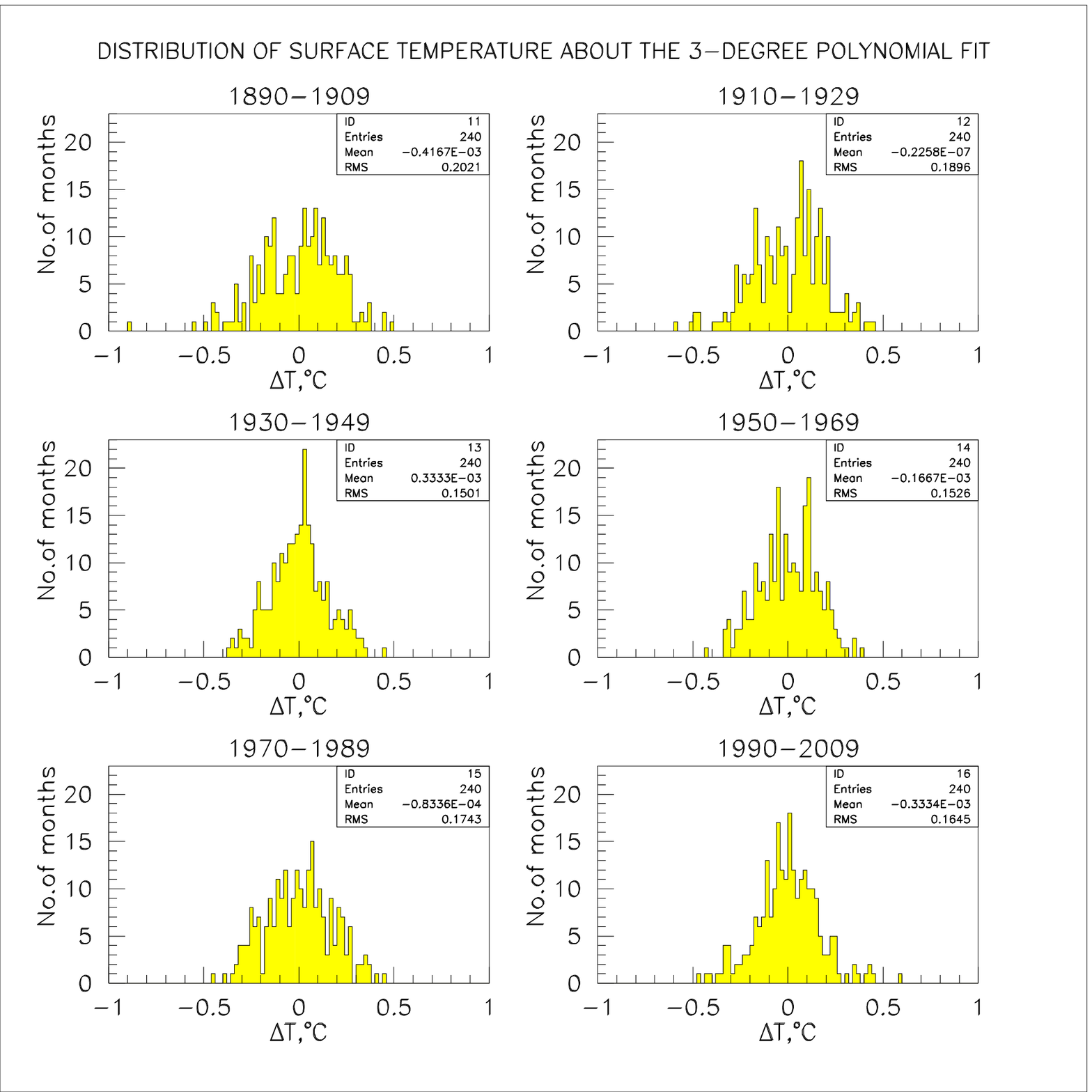}
\caption{}
\label{fig03}
\end{center}
\end{figure}

\newpage
\begin{figure}[ht]
\begin{center}
\includegraphics[height=15cm,width=15cm,angle=-90]{fluc_fig04.eps}
\caption{}
\label{fig04}
\end{center}
\end{figure}

\newpage
\begin{figure}[ht]
\begin{center}
\includegraphics[height=15cm,width=15cm]{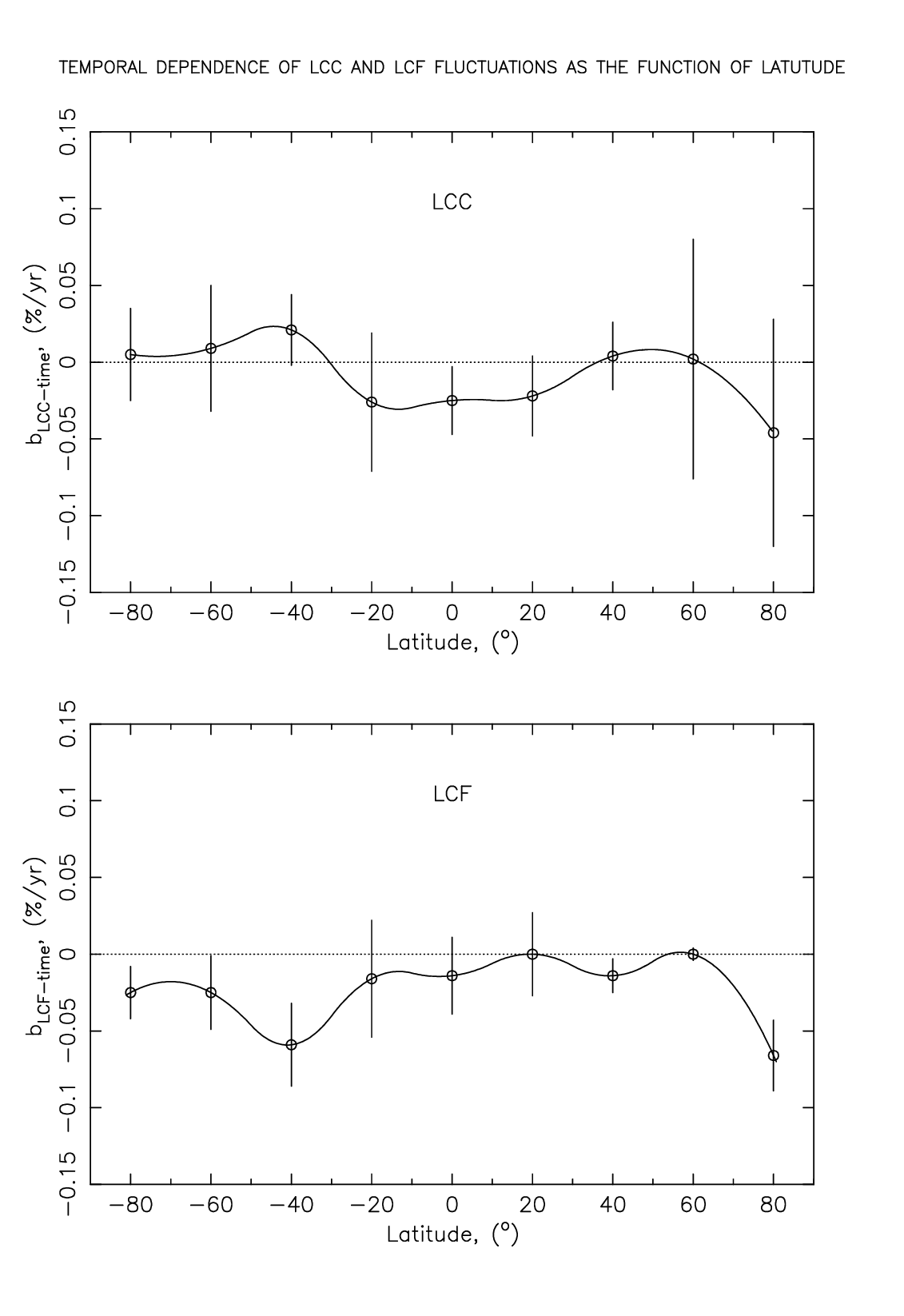}
\caption{}
\label{fig05}
\end{center}
\end{figure}

\end{document}